\documentclass[12pt]{article}
\usepackage[margin=1in]{geometry}
\usepackage{amsmath}
\usepackage{txfonts}

\usepackage{float}
\usepackage{booktabs}
\usepackage{indentfirst}
\usepackage{lipsum}
\usepackage{graphicx}
\usepackage{xcolor}
\usepackage{array}
\usepackage{enumitem}
\usepackage{hyperref}
\usepackage[utf8]{inputenc}
\DeclareUnicodeCharacter{202F}{ }
\hypersetup{
    colorlinks=true,
    linkcolor=blue,
    citecolor=blue,
    urlcolor=blue}

\newenvironment{keywords}{%
  \par\noindent\textbf{Keywords:}\ }{\par}
\title{Overtourism to Equilibrium: A System Dynamics \& Multi-Objective Model for Sustainable Destinations}
\author{%
\small
\begin{tabular}{>{\centering\arraybackslash}p{0.3\textwidth} >{\centering\arraybackslash}p{0.3\textwidth} >{\centering\arraybackslash}p{0.3\textwidth}}
\textbf{Huanzhu Lyu}\textsuperscript{1} & \textbf{Xiao Yang}\textsuperscript{2} & \textbf{Xintong Ji}\textsuperscript{3} \\
Computer Science & Engineering & Engineering \\
Central South University & Central South University & Central South University \\
Email: lvhuanzhuchat@gmail.com & Email: 2543648@dundee.ac.uk & Email: 15243687936@163.com
\end{tabular}
}
\date{}
\begin{document}
\maketitle
\begin{abstract}

Overtourism has posed severe challenges to many popular destinations worldwide, threatening both the natural environment and local communities. Aiming to balance \textbf{economic returns}, \textbf{environmental protection}, and \textbf{social satisfaction}, this paper develops a comprehensive decision-making model that integrates \textbf{system dynamics} with a \textbf{multi-objective evolutionary algorithm} (specifically, \textbf{NSGA-II}). First, we collect multi-source data from 2008 to 2024—including annual visitor arrivals (up to 3.1 million), government revenue and expenditure (up to \(\$1.03 \times 10^7\)), glacier retreat (220--350 ft), CO\(_2\) emissions (77,000--104,800 tons), and social satisfaction (ranging from 0.48 to 0.29)—and establish a dynamic system comprising four modules: tourist behavior, government finance, environmental evolution, and social well-being. These modules operate iteratively each year to simulate the evolution of the tourism system.

Next, we select critical decision variables such as the tax rate (\textit{tax\_rate}), carbon fee (\textit{carbon\_fee}), capacity limit (\textit{capacity\_limit}), and proportions of extra revenue allocated (e.g., \(\theta_{\text{env}}\), \(\theta_{\text{community}}\)), then apply \textbf{NSGA-II} to optimize three objectives: cumulative net revenue, final environmental index, and final social satisfaction. Experiments on Juneau’s dataset show that the optimal solution can yield cumulative net revenue up to \(\$1.64 \times 10^9\) with an environmental index of about 0.93 and social satisfaction at 0.86. If community-oriented policies are prioritized, social satisfaction can rise to 0.92 while net revenue drops to roughly \(\$1.05 \times 10^9\). Extending the model to Iceland—factoring in international flight volumes (reaching 3.2 million arrivals), a Ring Road capacity up to \(5 \times 10^6\), and distinct environmental and social characteristics—reveals a diverse Pareto front spanning net revenues of \(\$1.5 \times 10^8\)--\(\$2.0 \times 10^8\), environment indices up to 0.92, and social satisfaction levels above 0.80.

To further validate model robustness, we incorporate \textbf{Sobol} and \textbf{Morris} analyses. Results indicate that carbon fees and price elasticity account for over 60\% of the explanatory power on final environmental outcomes, whereas capacity limits explain around 90\% of net revenue variability. This underscores the importance of accurately gauging visitor price sensitivity and strategically scaling infrastructure to balance ecological preservation with economic growth. Finally, scenario simulations and flow-control strategies demonstrate how imposing capacity or dynamic pricing on crowded attractions—while directing more marketing and infrastructure investment to lesser-known sites—effectively mitigates congestion, raises social satisfaction, and enhances overall environmental measures. Our findings offer quantitative support for policymakers and tourism administrators, demonstrating that a multi-objective, system-dynamics model can be flexibly applied to address overtourism in diverse regions.

In summary, this work makes three main contributions: (i) it proposes an integrated system-dynamics and NSGA-II framework for sustainable tourism management that jointly optimizes economic, environmental, and social objectives; (ii) it demonstrates the framework’s portability via detailed case studies on Juneau and Iceland, including scenario-based policy analysis; and (iii) it provides a global sensitivity analysis (Morris and Sobol) that highlights the most influential policy levers and parameters, offering interpretable guidance for real-world decision makers.

\end{abstract}
\section{Introduction}
\subsection{Problem Background}
As increasingly more tourists flock into Juneau, the environments and residents have been affected profoundly. The tourists bring in significant revenue, but also overcrowding and more carbon footprint. We can see the glacier situation in Juneau from the figure~\ref{fig:Glacial Retreat Visualization} below:

\begin{figure}[h]
    \centering
    \includegraphics[width=14cm]{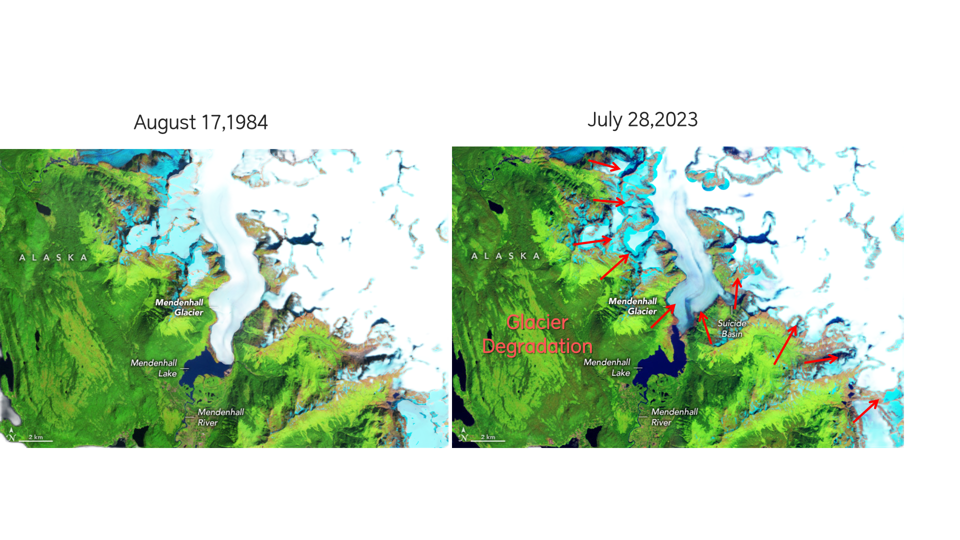}
    \caption{Glacial Retreat Visualization}
    \label{fig:Glacial Retreat Visualization}
\end{figure}

By comparison, we can see that the glacier has receded over time and experienced significant loss \cite{Davies2022}. There are growing concerns that the tourism industry in Juneau may eventually disappear with the glacier \cite{Eijgelaar2010}. In response to the situation above, establishing a mathematical model to efficiently deal with overtourism cannot be ignored anymore.
\subsection{Restatement of the Problem}

Our task is to promote the sustainable development of the tourism industry in Juneau, Alaska.  Following the MCM/ICM contest requirements, we address this complex issue by breaking it down into three key components:

\begin{enumerate}
    \item \textbf{Maximizing Tourism Benefits and Managing Uncertainty:} We develop a model to maximize the economic and social benefits of tourism in Juneau while considering inherent uncertainties. This involves creating a comprehensive revenue and expenditure plan, incorporating feedback loop mechanisms to adapt to changing conditions, and conducting sensitivity analysis to identify key factors influencing the model's outcomes.
    \item \textbf{Model Portability and Adaptability:}  Recognizing that overtourism affects numerous destinations, we focus on designing a portable and adaptable model.  This ensures its applicability to other locations facing similar challenges, maximizing its impact and promoting sustainable tourism practices beyond Juneau.
    \item \textbf{Balancing Tourist Distribution and Promoting Less-Visited Attractions:} To mitigate the negative impacts of concentrated tourist flows, we aim to balance tourist distribution. Our model explores strategies to promote less-visited attractions, creating a more diverse and enriching experience for visitors while relieving pressure on popular sites.  This approach fosters a more sustainable and enjoyable tourism ecosystem for both visitors and residents.
\end{enumerate}

\subsection{Literature Review}
Juneau is a major glacier tourism destination in Alaska, but its most famous tourist attraction—Mendenhall Glacier—has been experiencing accelerated melting in recent years, with data showing that the snowfall rate in the Juneau Icefield is nearly five times faster than previously recorded\cite{Borenstein2024}. Research indicates that additional greenhouse gas emissions from tourism are a significant factor in exacerbating glacier issues\cite{Bohrer2023}. The growing tourism activities (such as tourist transportation, accommodation operations, etc.) contribute to the deterioration of the scenic area's environment, while the development trend of local tourism depends on various factors including environmental conditions.

This complex causal relationship highlights the advantages of system dynamics in characterizing environmental–economic–social coupled systems\cite{Sedarati2018,Pooyan2018}. Meanwhile, multi-objective optimization methods (e.g., NSGA-II) are widely used to find optimal balance points among multiple dimensions including economic benefits, environmental protection, and social welfare, providing a scientific basis for policy-making. Existing studies, however, often \textit{either} focus on qualitative system-dynamics insights without explicit multi-objective optimization, \textit{or} emphasize optimization while simplifying feedback loops and sensitivity analysis. Beyond tourism, these methodological frameworks are general and can be used in urban planning, resource management, and industrial structure adjustment, but there remains a need for integrated models that combine realistic feedback structures, multi-objective search, and global sensitivity analysis in one coherent decision-support tool.

\subsection{Our Work}
\begin{figure}[H]
    \centering
    \includegraphics[width=16cm]{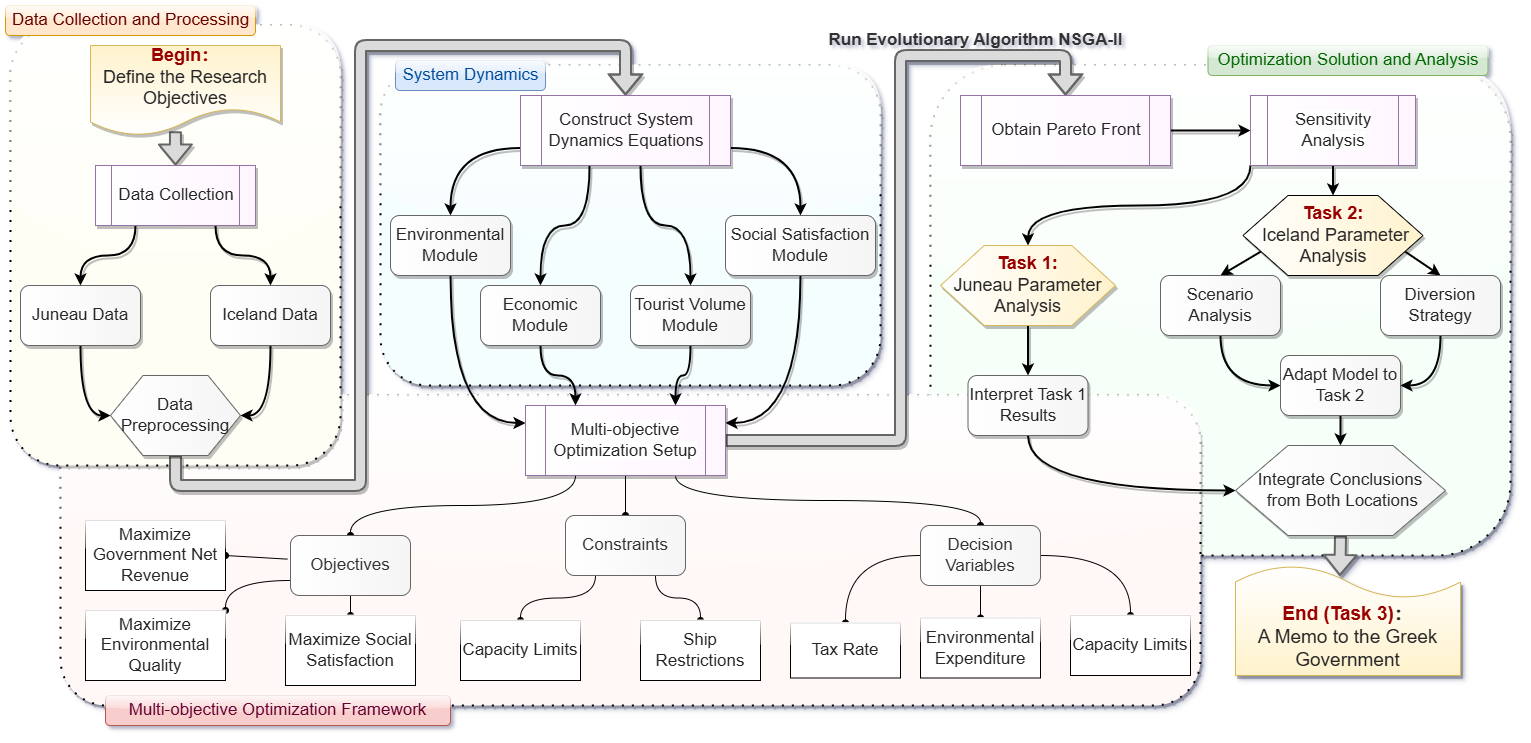}
    \caption{The Framework of Our Work}
    \label{fig:ourworkflowchart}
\end{figure}

\subsection{Contributions}
Building on the above background and literature, this paper makes the following contributions:
\begin{itemize}
    \item \textbf{Integrated framework}: We construct an integrated Juneau Sustainable Tourism (JST) model that couples system dynamics with a multi-objective evolutionary algorithm (NSGA-II), enabling simultaneous optimization of cumulative net revenue, environmental quality, and social satisfaction under realistic feedback structures.
    \item \textbf{Portable case studies}: We calibrate and apply the framework to two representative destinations—Juneau and Iceland—demonstrating how the same modeling and optimization pipeline can be adapted to different data, capacity constraints, and policy contexts.
    \item \textbf{Global sensitivity and scenario analysis}: We perform Morris and Sobol global sensitivity analyses together with scenario simulations to identify key parameters (e.g., price elasticity, carbon fees, capacity limits) and to visualise policy trade-offs, thereby providing interpretable guidance for decision makers.
\end{itemize}

This work originated from our entry to the 2025 MCM/ICM contest, where it received the Meritorious Winner (M) award. Here, we present a refined and extended version aimed at the broader research and policy-making community, positioning the model as a generalizable computational decision-support tool for sustainable tourism.

\section{Model Preparation}

\subsection{Assumptions and Justifications}

\textbf{Assumption 1:} We assume there are no major natural disasters (i.e., epidemics) or dramatic policy changes. 

\textit{Justification:} Epidemics have a huge impact on tourism, with missing tourist data during such periods, and major events are difficult to predict.

\textbf{Assumption 2:} Time is discretized into years, with variables in the model evolving and updating between years. 

\textit{Justification:} Tourism has seasonality, with factors like visitor numbers showing annual periodic changes; additionally, main statistical data and policy-making are conducted on an annual basis.

\textbf{Assumption 3:} Glacier melting is quantifiable, and if glaciers retreat too quickly, the "glacier attractiveness" in subsequent years will decrease, affecting tourists' willingness to visit.

\textit{Justification:} Converting abstract glacier changes into specific numerical indicators is necessary for them to serve as decision variables in the model.

\textbf{Assumption 4:} Carbon emissions are assumed to be mainly contributed by tourism-related activities, with emission levels positively correlated with tourist numbers. 

\textit{Justification:} Tourism-related transportation, accommodation, and visiting activities, all with limited scale, account for carbon emissions of over 90\%. It is the actual characteristic of the system and simplifies the model.

\textbf{Assumption 5:} The effect of exchange rate fluctuations is not taken into account. 

\textit{Justification:} Tourism in Juneau is mainly a domestic U.S. market, cruise package prices are mostly in USD, and it accounts for a small proportion (<15\%) of total tourism.

\textbf{Assumption 6:} Social satisfaction is assumed to be mainly influenced by three factors: local economic benefits, negative externalities (such as congestion, housing prices, noise, etc.), and environmental quality. 

\textit{Justification:} The impact of these three factors comprehensively reflects the perception of quality of life by residents and jointly determines residents' living standards and environmental comfort.

\subsection{Notations}

\begin{table}[H]
\centering
\begin{tabular}{cccc}
\toprule
\textbf{Symbol} & \textbf{Description} & \textbf{Symbol} & \textbf{Description} \\
\midrule
$\tau$ & Tourism tax rate & $\rho_{\text{env}}$ & Environmental spending ratio \\
$\delta_{\text{dev}}$ & Development incentive & $C_{\max}$ & Tourist capacity limit \\
$S_{\max}$ & Annual ship limit & $\phi_{\text{carbon}}$ & Carbon emission fee \\
$\rho_{\text{glacier}}$ & Glacier fund ratio & $V_t$ & Total visitors in year $t$ \\
$E_t$ & Environmental index & $S_{\text{sat},t}$ & Social satisfaction \\
$R_{\text{net\_cum},t}$ & Cumulative net revenue & $V_{\text{base}, t}$ & Base visitors \\
$R_{\text{gov\_base}, t}$ & Base government revenue & $\text{EXP}_{\text{gov\_base}, t}$ & Base government expenditure \\
$G_{\text{retreat}, t}$ & Glacier retreat amount & $\text{CO2}_{\text{emission}, t}$ & CO$_2$ emissions \\
$S_{\text{sat\_base}, t}$ & Base social satisfaction & $P_t$ & Population \\
$U_t$ & Unemployment rate & $P_{\text{visitor\_base}}$ & Base ticket price \\
$\varepsilon_{\text{price}}$ & Price elasticity coefficient & $G_{\text{retreat\_baseline}}$ & Baseline glacier retreat \\
$\kappa$ & Glacier retreat impact coefficient & $P_{\text{ship\_capacity}}$ & Average ship capacity \\
$\alpha_g$ & Glacier protection coefficient & $\alpha_w$ & Waste treatment coefficient \\
$\beta_1$ & Glacier retreat impact & $\beta_2$ & CO$_2$ emission impact \\
$\delta$ & Natural recovery coefficient & $p_{\text{glacier}}$ & Glacier protection satisfaction \\
$p_{\text{waste}}$ & Waste treatment satisfaction & $p_2$ & Crowding impact \\
$p_3$ & Environmental quality impact & $p_4$ & Unemployment impact \\
\bottomrule
\end{tabular}
\end{table}
\textbf{Note:} There are some notations that are not listed here and will be discussed in detail in each section.

\subsection{Data Pre-processing}

Data for each year (2008–2024) regarding visitor arrivals, government revenue/expenditure, glacier retreat, CO\textsubscript{2} emissions, population, and social satisfaction are first compiled into a unified table. Missing or partial records are then addressed by linear interpolation (or forward/backward filling when appropriate) to ensure continuous time series. Finally, each variable is checked for outliers against known trends (e.g., plausible ranges of glacier retreat) and normalized if necessary, producing a consistent dataset ready for system dynamics simulation and multi-objective optimization.

\section{JST Model Construction}
To better address sustainable tourism management in the Juneau area, we plan to combine System Dynamics (SD) with multi-objective optimization (evolutionary algorithms as NSGA-II) to construct a dynamic model that can simultaneously focus on the triple objectives of "economy-environment-society".

Juneau Sustainable Tourism model consists of four major sub-modules: tourist demand and capacity, government finance, environmental evolution, and social satisfaction. We have established a series of adjustable decision variables, then used System Dynamics to build sub-models to simulate their dynamic impacts on the aforementioned objectives (economy-environment-society) over several years. Finally, we employ evolutionary algorithms (genetic algorithms, NSGA-II) for multi-objective optimization of these three objectives to obtain Pareto Frontier solutions that are relatively optimal across different objective dimensions.

The following sections will introduce the establishment and core equations of each sub-module in sequence, and briefly summarize the iteration process of the entire system and the solution approach for multi-objective optimization.

\subsection{Model I: System Dynamics Model }
In the tourism management of Juneau, due to countless interactions between various factors related to tourism and environmental protection, it can be viewed as a complex system. System Dynamics can catch the subsystem interactions effectively, feedback of policy simulation, and long-term dynamics due to multiple variables \cite{Pooyan2018}.

\subsubsection{System Boundary Definition and Structural Analysis}
Considering that glacier attractions are a major feature of Juneau's tourism, this study defines a system boundary that includes the following four main subsystems:

\begin{itemize}
    \item \textbf{Environmental Subsystem (G):} Includes impacts from climate change and greenhouse gas emissions.
    \item \textbf{Tourism Subsystem (T):} Includes impacts from tourism activities.
    \item \textbf{Social Subsystem (S):} Includes impacts from community surveys and local resident income.
    \item \textbf{Economic Subsystem (B):} Includes government revenue (taxes, investments) and expenditure (environmental protection construction, baseline spending).
\end{itemize}

Economic Subsystem: Includes government revenue (taxes, investments) and expenditure (environmental protection construction, baseline spending) 

\begin{figure}[h]
    \centering
    \includegraphics[width=8cm]{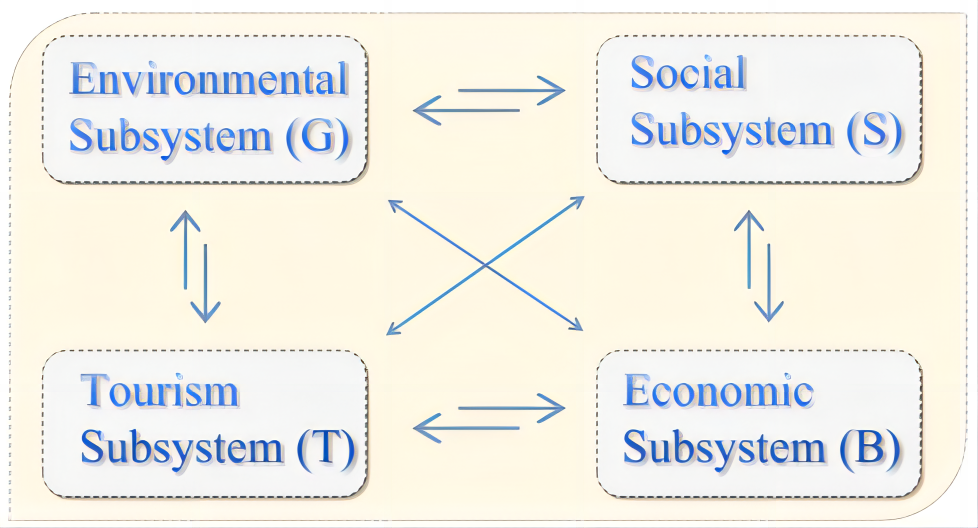}
    \caption{Relationship Diagram Between Subsystems}
    \label{fig:subsystems}
\end{figure}

The Environmental Subsystem (G) centers on the Mendenhall Glacier, the Tourism Subsystem (T) focuses on tourist numbers, the Social Subsystem (S) centers on local resident satisfaction, and the Economic Subsystem (B) reflects government finance and investment.
As shown in Figure~\ref{fig:subsystems}, these subsystems interact and are inseparably interconnected, forming a dynamically balanced overall system.

\subsubsection{Causal Relationship Analysis}

\begin{figure}[H]
    \centering
    \includegraphics[width=16cm]{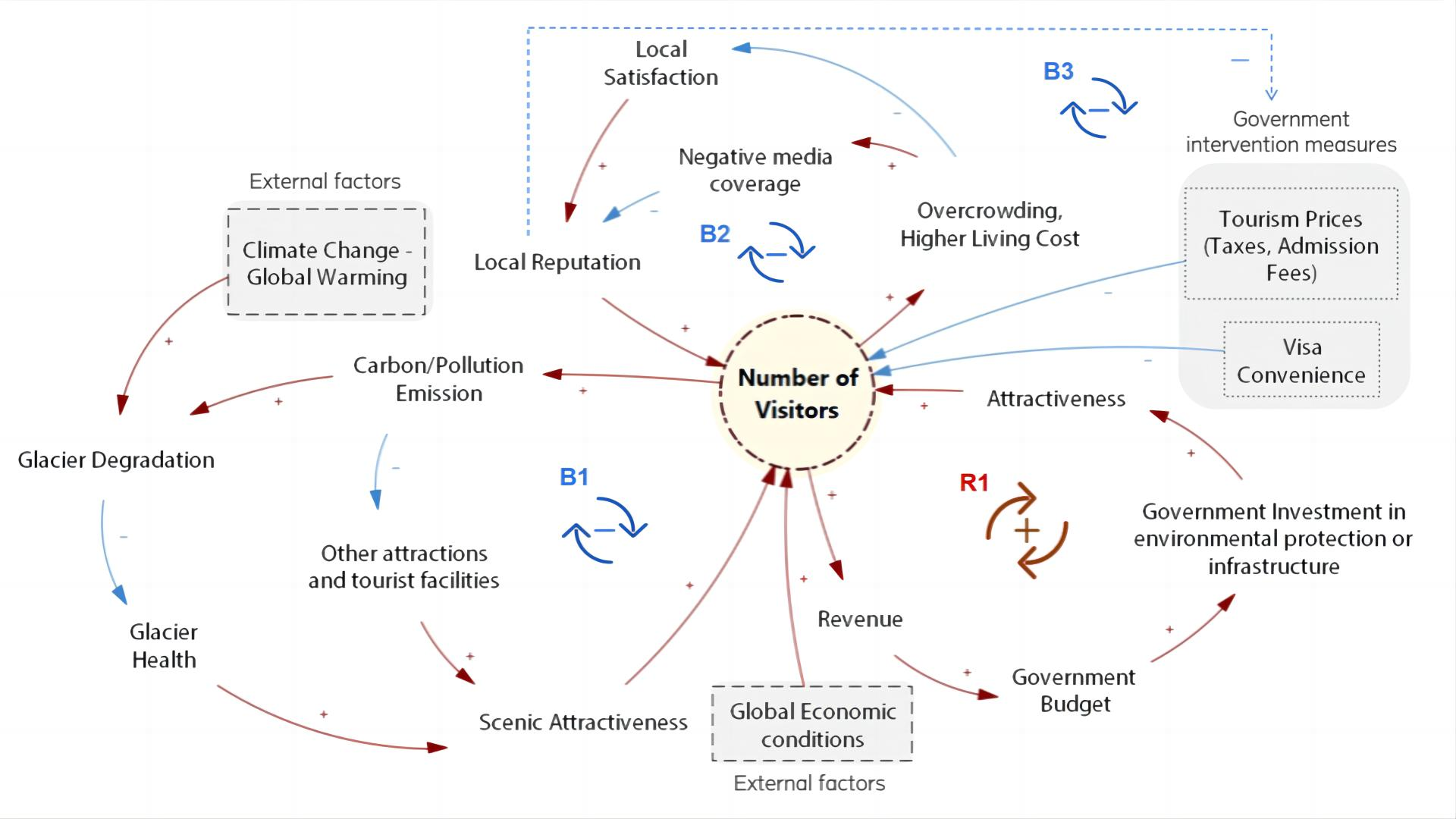}
    \caption{Causal Loop Diagram}
    \label{fig:Causal Loop Diagram}
\end{figure}

The causal loop diagram as shown in Figure~\ref{fig:Causal Loop Diagram} is based on 'Number of Tourists' and reflects the complex relationships between different pieces in the Juneau tourism system. The diagram includes four major feedback loops: one reinforcing loop R1 (formed through income-government budget-infrastructure investment) and three balancing loops B1, B2, and B3 (formed through scenic attractiveness, local reputation, and government intervention respectively). The system is affected from external factors like global economic conditions, climate weather changes that in turn affect glacier health and local satisfaction factors with relation to carbon emissions and the crowding levels. By adjusting tourism prices and the convenience of visa, the government can balance system development. 

\subsubsection{Dynamic System Modeling}
\begin{figure}[h]
    \centering
    \includegraphics[width=15cm]{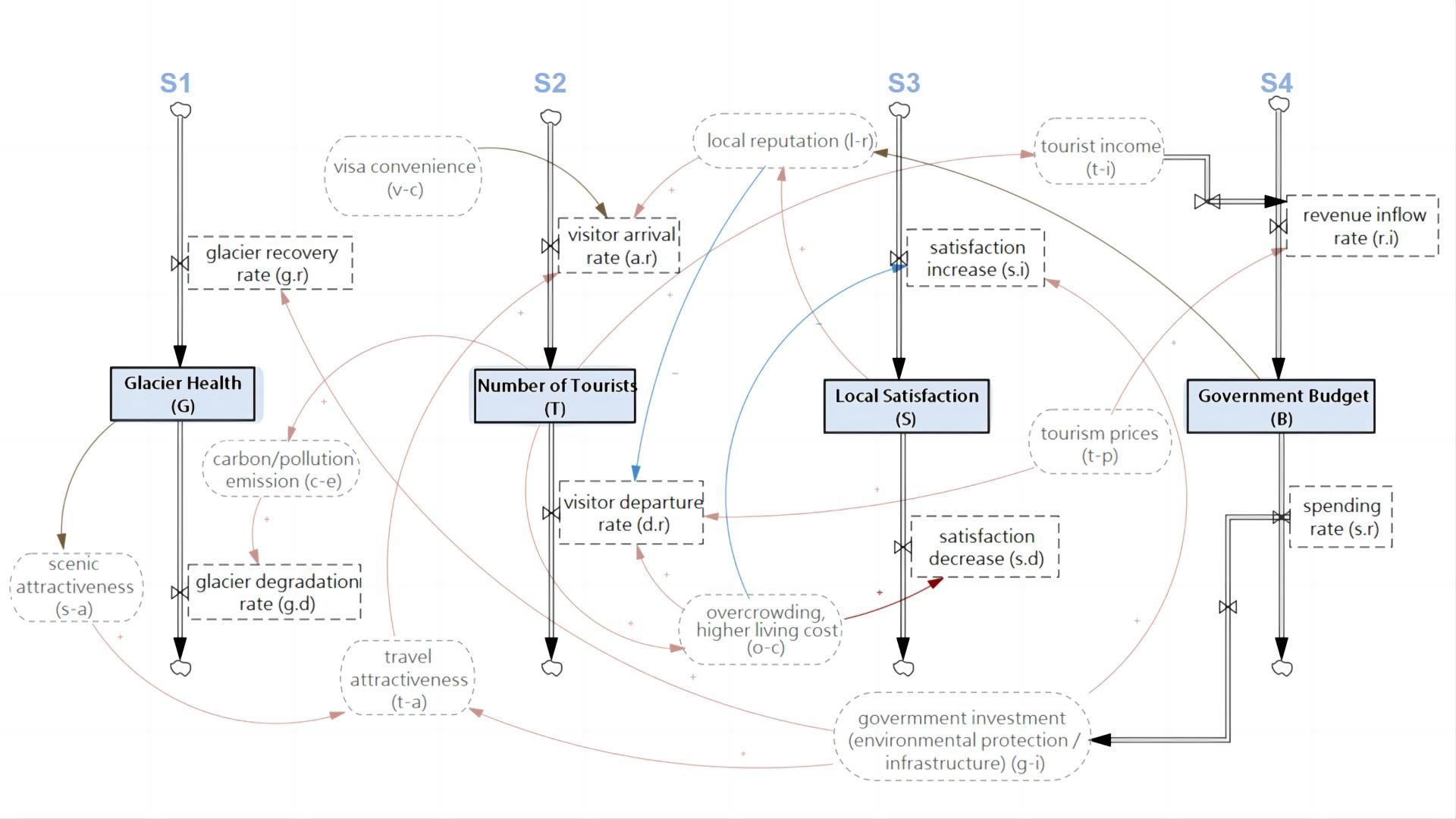}
    \caption{Stock-Flow Diagram}
    \label{fig:Stock-Flow Diagram}
\end{figure}
As shown in the Figure~\ref{fig:Stock-Flow Diagram}, subsystems form interconnected relationships through feedback loops of auxiliary and flow variables, collectively determining the sustainable tourism development state of Juneau.

\begin{enumerate}[label=(\alph*)]
\item Tourist numbers depend on the scenic appeal of glacier health; conversely, tourist activities are the source of carbon emissions, which affect glacier health.

\item The levels of crowding affect tourist's satisfaction and it further affects local reputation which eventually affects tourism.

\item Tourist numbers increase income from tourism, which the government budget for infrastructure investment helps to increase tourism attractiveness.

\item Government investment affects local satisfaction; while local satisfaction indirectly influences tourism income through reputation.
\end{enumerate}

Next, we will construct corresponding sub-models based on the stock-flow relationships of each subsystem.

\subsubsection{Model Equation Design}
Following the causal loop and stock-flow structures described above, we now present the specific equations that capture the dynamics of each subsystem. These equations are designed to reflect the reinforcing and balancing feedback loops among tourist numbers, government finances, local satisfaction, and environmental conditions. In particular, they align with the idea that tourist demand in Juneau depends on factors like glacier health and crowding, while government interventions and local attitudes feed back to shape both future visitation and the region’s sustainability.

\textbf{Visitor Dynamics}  
Tourist numbers evolve each year through an interplay of effective tourism price, glacier-based scenic attractiveness, capacity constraints, and potential social resistance. The effective tourism price \(P_{\text{effective}, t}\) is defined as 
\begin{equation}
P_{\text{effective}, t} = \bigl(P_{\text{visitor\_base}} + \varphi_{\text{carbon}}\bigr)\,(1 + \tau), \label{eq:effective_price}
\end{equation}
where \(P_{\text{visitor\_base}}\) is the baseline ticket price, \(\varphi_{\text{carbon}}\) is a per-visitor carbon fee, and \(\tau\) is the tourism tax rate. A higher \(\tau\) or carbon fee increases the overall cost for visitors. The glacier attractiveness factor 
\begin{equation}
F_{\text{glacier}, t} = \max\bigl(0,\,1 \;-\; \kappa\,\bigl(\tfrac{G_{\text{retreat}, t}}{G_{\text{retreat\_baseline}}} - 1\bigr)\bigr) \label{eq:glacier_factor}
\end{equation}
captures how glacier retreat reduces scenic appeal, modulated by the sensitivity parameter \(\kappa\). When the glacier recedes significantly beyond a baseline level, this factor diminishes toward zero. A composite measure of attraction 
\begin{equation}
F_{\text{attraction}, t} = \bigl(1 + \alpha\,(E_t + S_{\text{sat}, t} - 1)\bigr)\,F_{\text{glacier}, t} \label{eq:attraction_factor}
\end{equation}
then incorporates both environmental quality \(E_t\) and social satisfaction \(S_{\text{sat}, t}\) to further reflect the tourism draw. The parameter \(\alpha\) is set to 0.5.
To reflect development incentives, we introduce an exogenous boost in visitors:
\begin{equation}
\Delta V_{\text{dev}, t} \;=\; \delta_{\text{dev}} \;\times\; K_{\text{dev}}, \label{eq:dev_incentive}
\end{equation}
where \(\delta_{\text{dev}}\) is a policy-controlled parameter and \(K_{\text{dev}}\) is a constant, representing the approximate annual increase in tourists from moderate marketing or infrastructure initiatives. The price elasticity factor 
\begin{equation}
F_{\text{price}, t} = 1 + \varepsilon_{\text{price}}\,\bigl(\tau + \tfrac{\varphi_{\text{carbon}}}{k_1}\bigr) \label{eq:price_factor}
\end{equation}
accounts for how visitors react to changes in taxes and fees. The parameter \(k_1\) is set to 100. If \(\varepsilon_{\text{price}}\) is negative, an increase in \(\tau\) or \(\varphi_{\text{carbon}}\) lowers demand. Combining these effects, the potential (unconstrained) visitor number in the next period is
\begin{equation}
V'_{t+1} = V_{\text{base},\,t+1}\,F_{\text{price}, t}\,F_{\text{attraction}, t} + \Delta V_{\text{dev}, t}, \label{eq:potential_visitors}
\end{equation}
where \(V_{\text{base},\,t+1}\) is the baseline forecast. Actual arrivals must then obey capacity and vessel constraints through
\begin{equation}
V_{t+1} = \min\bigl(V'_{t+1},\,C_{\max},\,S_{\max}\,P_{\text{ship\_capacity}}\bigr), \label{eq:actual_visitors}
\end{equation}
limiting growth when the expected number exceeds the city’s accommodation or the allowed number of cruise ships \((S_{\max})\). If local residents become strongly dissatisfied \(S_{\text{sat}, t} < S_{\text{threshold}}\), a social resistance effect cuts arrivals by a fixed ratio:
\begin{equation}
\text{if } S_{\text{sat}, t}<S_{\text{threshold}}:\; V_{t+1} \leftarrow V_{t+1}\,\times\, R_{\text{social}}, \label{eq:social_resistance}
\end{equation}
This discourages tourism in times of community pushback. The parameter \(S_{\text{threshold}}\) is set to 0.3 and \(R_{\text{social}}\) is set to 0.8.

\textbf{Government Finance Dynamics}  
Government revenue accumulates from baseline revenue, tourism taxes, carbon fees, and any development funds, while expenditures cover baseline costs and environmental efforts. Tourism-based revenue is 
\begin{equation}
R_{\text{tourism}, t} = V_{t+1}\,\bigl(P_{\text{visitor\_base}}\tau \;+\; \varphi_{\text{carbon}}\bigr), \label{eq:tourism_revenue}
\end{equation}
scaled by the actual visitor count \(V_{t+1}\). The total government income 
\begin{equation}
R_{\text{gov\_total}, t} = R_{\text{gov\_base}, t} + R_{\text{tourism}, t} + \delta_{\text{dev}}\,\times\,K_{\text{gov\_dev}} \label{eq:total_gov_revenue}
\end{equation}
combines baseline finances \(R_{\text{gov\_base}, t}\), tourism revenue, and dedicated development funds. The parameter \(K_{\text{gov\_dev}}\) is set to \(10^5\). A fixed fraction \(\rho_{\text{env}}\) of this sum is allocated to environmental protection,
\begin{equation}
\text{EXP}_{\text{env}, t} = \rho_{\text{env}}\,R_{\text{gov\_total}, t}. \label{eq:env_spending}
\end{equation}
Total public spending 
\begin{equation}
\text{EXP}_{\text{gov\_total}, t} = \alpha_{\text{gov\_base}}\,\text{EXP}_{\text{gov\_base}, t} + \text{EXP}_{\text{env}, t} \label{eq:total_gov_spending}
\end{equation}
assumes that only a fraction \(\alpha_{\text{gov\_base}}\) of the baseline outlays are closely linked to tourism, with the rest considered inflexible. The parameter \(\alpha_{\text{gov\_base}}\) is set to 0.3. The net annual surplus 
\begin{equation}
R_{\text{net}, t} = R_{\text{gov\_total}, t} - \text{EXP}_{\text{gov\_total}, t} \label{eq:net_revenue}
\end{equation}
shows whether the city’s tourism-related accounts are profitable. Summing it over the years yields a cumulative figure,
\begin{equation}
R_{\text{net\_cum}, t+1} = R_{\text{net\_cum}, t} + R_{\text{net}, t}, \label{eq:cumulative_revenue}
\end{equation}
which monitors longer-term fiscal stability.

\textbf{Environmental Quality Dynamics}  
Environmental quality in Juneau is tracked by an index \(E_t \in [0,1]\). Investments in glacier protection and waste treatment have a positive impact on \(E_t\), while glacier retreat, CO\(_2\) emissions, and crowding exert negative pressures. Of the total environmental budget \(\text{EXP}_{\text{env}, t}\), a fraction \(\rho_{\text{glacier}}\) is devoted to glacier protection:
\begin{equation}
\text{EXP}_{\text{glacier}, t} = \rho_{\text{glacier}}\;\text{EXP}_{\text{env}, t}, \quad \text{EXP}_{\text{waste}, t} = (1-\rho_{\text{glacier}})\,\text{EXP}_{\text{env}, t}. \label{eq:env_spending_allocation}
\end{equation}
The amounts allocated to glacier and waste initiatives contribute positively as
\begin{equation}
\Delta E_{\text{spend\_glacier}, t} = \alpha_g\,\text{EXP}_{\text{glacier}, t}\,(1 - E_t),\quad \Delta E_{\text{spend\_waste}, t} = \alpha_w\,\text{EXP}_{\text{waste}, t}\,(1 - E_t), \label{eq:env_spending_impact}
\end{equation}
where \(\alpha_g\) and \(\alpha_w\) represent effectiveness coefficients that produce a stronger effect when \(E_t\) is relatively low. Meanwhile, glacier recession and CO\(_2\) drive down the index,
\begin{equation}
\Delta E_{\text{retreat}, t} = -\beta_1\,G_{\text{retreat}, t}, \quad \Delta E_{\text{CO2}, t} = -\beta_2\,\text{CO2}_{\text{emission}, t}. \label{eq:env_negative_impact}
\end{equation}
A natural regeneration term 
\begin{equation}
\Delta E_{\text{recover}, t} = \delta\,(1 - E_t) \label{eq:env_recovery}
\end{equation}
prevents the index from dropping irreversibly, reflecting nature’s capacity to restore itself over time. Summing these contributions updates the index,
\begin{equation}
E_{t+1} = E_t + \Delta E_{\text{spend\_glacier}, t} + \Delta E_{\text{spend\_waste}, t} + \Delta E_{\text{retreat}, t} + \Delta E_{\text{CO2}, t} + \Delta E_{\text{recover}, t}, \label{eq:env_index_update}
\end{equation}
and \(E_{t+1}\) is bounded within \([\,0, 1\,]\) to maintain interpretability.

\textbf{Social Satisfaction Dynamics}  
Local satisfaction \(S_{\text{sat}, t}\) also ranges from 0 to 1, influenced by environmental investment, crowding, unemployment, and the environment itself. Higher government spending on glacier protection and waste management raises community morale:
\begin{equation}
\Delta S_{\text{spend\_glacier}, t} = p_{\text{glacier}}\,\text{EXP}_{\text{glacier}, t}\,(1 - S_{\text{sat}, t}),\quad \Delta S_{\text{spend\_waste}, t} = p_{\text{waste}}\,\text{EXP}_{\text{waste}, t}\,(1 - S_{\text{sat}, t}), \label{eq:social_spending_impact}
\end{equation}
with \(p_{\text{glacier}}\) and \(p_{\text{waste}}\) capturing the boost from conservation efforts. Crowding and unemployment lower satisfaction through
\begin{equation}
\Delta S_{\text{crowd}, t} = -\,p_2\,\tfrac{V_{t+1}}{P_t + \varepsilon}, \quad \Delta S_{\text{unemp}, t} = -\,p_4\,U_t, \label{eq:social_negative_impact}
\end{equation}
where \(V_{t+1}\!/P_t\) approximates perceived congestion, and \(U_t\) is the unemployment rate. Environmental quality also feeds back into community sentiment:
\begin{equation}
\Delta S_{\text{env}, t} = p_3\,(E_{t+1} - S_{\text{sat}, t}), \label{eq:social_env_impact}
\end{equation}
so a healthier environment nudges satisfaction upward. Collectively,
\begin{equation}
S_{\text{sat}, t+1} = S_{\text{sat}, t} + \Delta S_{\text{spend\_glacier}, t} + \Delta S_{\text{spend\_waste}, t} + \Delta S_{\text{crowd}, t} + \Delta S_{\text{unemp}, t} + \Delta S_{\text{env}, t} \label{eq:social_satisfaction_update}
\end{equation}
defines the next period’s community satisfaction, constrained again within \([\,0,1\,]\).

By translating the causal loop and stock-flow diagram into these equations, we formalize how each subsystem evolves year by year and how their interactions reinforce or balance each other. Tourist numbers feed into government revenue, which supports environmental and social investment, while environmental conditions and local attitudes feed back to regulate tourism flow. This model thus captures the multifaceted feedback loops (R1, B1, B2, B3) outlined above and provides a dynamic basis for evaluating sustainable tourism policies in Juneau.

\subsection{Model II: Multi-Objection Optimization Model}
Due to the interactions, constraints, conflicts, and trade-offs between the three objectives of "economy-environment-society," this research aims to find Pareto-optimal solutions through multi-objective optimization for Juneau's tourism development system. In this way, we aim to attain the best economical benefits and environmental protection as well as social well being simultaneously and offer decision makers diversified options.

\subsubsection{Constraints on Decision Variables and Multi-objective Optimization Functions}
In this research, we select the maximization of net income, environmental health, and social satisfaction as the objective functions for multi-objective optimization based on sustainable tourism development requirements. These three objective functions not only provide scientific decision support for specific policy formulation and implementation but also reflect the essential demands of sustainable development.

Based on the above analysis, we established system dynamics equations that describe how key state variables (e.g., \(V_t\), \(E_t\), \(S_{sat,t}\), \(R_{net\_cum,t}\)) evolve over time. Through these equations, we construct a multi-objective optimization model aimed at simultaneously maximizing cumulative net revenue, final environmental quality, and social satisfaction by selecting appropriate decision variables.

\textbf{(1) Objective Functions}

Based on our analysis of the practical background and understanding of the problem context, sustainable tourism development requires balancing economic viability, environmental protection, and social equity. Concretely, we define three objectives evaluated at the final time \(T\):

\begin{enumerate}
    \item \textbf{Maximize Cumulative Net Revenue}
    \begin{equation}
    \max f_1 \;=\; R_{\text{net\_cum},\,T}.
    \end{equation}
    Here, \(R_{\text{net\_cum},\,T}\) represents the \textbf{accumulated net income} from tourism and related activities over the entire simulation horizon \([1,\dots,T]\). A higher \(f_1\) implies stronger fiscal capacity for the government, enabling further public investments and resilience against economic shocks.

    \item \textbf{Maximize Final Environmental Index}
    \begin{equation}
    \max f_2 \;=\; E_T.
    \end{equation}
     The variable \(E_T \in [0,1]\) summarizes overall \textbf{environmental quality} at the end of the planning period—incorporating glacier health, waste management outcomes, and natural regeneration. Achieving a higher \(E_T\) indicates that Juneau’s delicate ecosystem is safeguarded or improved, ensuring long-term sustainability.

    \item \textbf{Maximize Final Social Satisfaction}
    \begin{equation}
    \max f_3 \;=\; S_{\text{sat},\,T}.
    \end{equation}
    Local residents’ well-being is captured by \(S_{\text{sat},\,T} \in [0,1]\). Prioritizing social satisfaction emphasizes inclusive development and aims to mitigate overtourism’s social costs, such as crowding and cultural erosion.
\end{enumerate}

Striving to \textbf{jointly} optimize these three goals reflects the core pillars of sustainable tourism and resonates with the balancing and reinforcing loops identified in the causal analysis.

\bigskip

\textbf{(2) Decision Variables}

To meet the above objectives, we control a set of \textbf{seven} decision variables:
\begin{equation*}
\bigl(\tau,\;\rho_{\text{env}},\;\delta_{\text{dev}},\;C_{\max},\;S_{\max},\;\varphi_{\text{carbon}},\;\rho_{\text{glacier}}\bigr).
\end{equation*}

Each variable influences one or more subsystems described in the system dynamics equations:

\begin{enumerate}
    \item \(\tau\): \textbf{Tourism tax rate} \([0, 0.3]\).
        \begin{itemize}
            \item Reflects the proportion of basic tourist spending or ticket price taxed by the local government. Empirical data suggest typical tourism taxes rarely exceed 30\%, so we set an upper bound of 0.3.
        \end{itemize}

    \item \(\rho_{\text{env}}\): \textbf{Fraction of total government revenue allocated to environmental protection} \([0, 0.5]\).
        \begin{itemize}
            \item To maintain flexibility for other budgetary needs, we cap this fraction at 50\%, based on observed practices in environmentally sensitive regions.
        \end{itemize}
    \item \(\delta_{\text{dev}}\): \textbf{Development incentive intensity} \([0, 1]\).
         \begin{itemize}
            \item A dimensionless parameter that scales additional visitor arrivals (\(\Delta V_{\text{dev},\,t} = \delta_{\text{dev}}\times K_{\text{dev}}\)) and potential government grants (\(\delta_{\text{dev}}\times K_{\text{fund}}\)) aimed at tourism growth. Ranging from 0 (no extra stimulation) to 1 (full policy push).
         \end{itemize}
    \item \(C_{\max}\): \textbf{Maximum accommodation capacity} \(\bigl[1\times10^6,\;4\times10^6\bigr]\).
        \begin{itemize}
             \item Encompasses lodging constraints, infrastructure, and daily handling capacity for visitors. The range reflects estimates from historic data on peak visitor loads and prospective expansions.
        \end{itemize}

    \item \(S_{\max}\): \textbf{Maximum number of cruise-ship (or flight) slots per period} \([600, 800]\).
        \begin{itemize}
             \item This represents how many large vessels the port or airport can safely handle over a season. Values derive from average daily/seasonal scheduling data, factoring potential expansions but also local environmental restrictions.
        \end{itemize}

    \item \(\varphi_{\text{carbon}}\): \textbf{Carbon fee (USD) per visitor} \([0, 100]\).
        \begin{itemize}
             \item Levied to internalize the externalities of greenhouse gas emissions. The upper bound of \$100/visitor is taken from carbon tax proposals in tourism-intense economies, ensuring the model can explore stricter environmental policies.
        \end{itemize}

    \item \(\rho_{\text{glacier}}\): \textbf{Proportion of environmental funds dedicated to glacier protection} \([0, 1]\).
        \begin{itemize}
           \item The remainder \((1 - \rho_{\text{glacier}})\) is assumed to go into waste management and other environmental measures. Since glacier health is a critical factor for Juneau’s allure, \(\rho_{\text{glacier}}\) can vary from 0 (focus on broader ecological investments) to 1 (all environment budget aimed at glacier conservation).
        \end{itemize}
\end{enumerate}
\bigskip

Consequently, solving this multi-objective problem means searching over \(\bigl(\tau,\;\rho_{\text{env}},\;\delta_{\text{dev}},\;\dots\bigr)\) within the above ranges and respecting the dynamic relationships. We then evaluate three \textbf{objective functions}—\(R_{\text{net\_cum},\,T}\), \(E_T\), and \(S_{\text{sat},\,T}\)—to identify “efficient frontiers” (Pareto-optimal solutions) that balance economic, environmental, and social dimensions of sustainable tourism in Juneau.

\subsubsection{NSGA-II-based Pareto Solution}

The overall process of NSGA-II is schematically shown in Figure~\ref{fig:NSGA-II} and involves the following steps:

\begin{figure}[h]
    \centering
    \includegraphics[width=15cm]{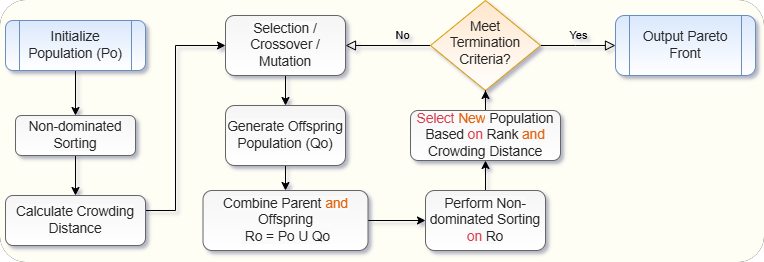}
    \caption{Flow Diagram of NSGA-II Algorithm}
    \label{fig:NSGA-II}
\end{figure}

\begin{enumerate}
    \item \textbf{Population Initialization}
    \begin{itemize}
        \item \textbf{Decision Variables}: Each individual (solution) in the population encodes the seven decision variables \(\bigl(\tau,\;\rho_{\text{env}},\;\delta_{\text{dev}},\;C_{\max},\;S_{\max},\;\varphi_{\text{carbon}},\;\rho_{\text{glacier}}\bigr)\).
        \item \textbf{Range Sampling}: We randomly initialize population of size \(N=100\) within the specified bounds. This ensures a broad initial coverage of the decision space (e.g., \(\tau\in[0,0.3]\), \(\rho_{\text{env}}\in[0,0.5]\), etc.).
    \end{itemize}

    \item \textbf{Fitness Evaluation}
    \begin{itemize}
        \item For each individual in the current population, we \textbf{simulate} the tourism system over \(T\) years using our system dynamics model.
        \item We then record its objective values:
        \[
        \bigl(F_1,\;F_2,\;F_3\bigr)
        \;=\;
        \Bigl(
        R_{\text{net\_cum},\,T},\;
        E_T,\;
        S_{\text{sat},\,T}
        \Bigr),
        \]
        corresponding respectively to cumulative net revenue, \textbf{final environmental index}, and final social satisfaction.
    \end{itemize}

    \item \textbf{Non-dominated Sorting}
    \begin{itemize}
         \item The population is sorted according to Pareto dominance: an individual \(A\) is said to dominate \(B\) if it is no worse in all objectives and strictly better in at least one.
        \item Solutions are assigned a Pareto rank (or front) based on how many solutions dominate them.
    \end{itemize}

    \item \textbf{Selection (Tournament + Crowding Distance)}
    \begin{itemize}
        \item Using \textbf{binary tournament selection} with non-dominated sorting ranks and \textbf{crowding distance} as tie-breakers, we pick high-quality individuals to become “parents” for genetic operators.
        \item This ensures that solutions from better Pareto fronts and with sparser neighborhoods in objective space have higher chances of reproduction.
    \end{itemize}

    \item \textbf{Crossover and Mutation}
    \begin{itemize}
        \item We apply crossover (SBX—simulated binary crossover) to exchange decision variable segments between selected parent pairs, producing new offspring solutions.
        \item We then mutate certain genes (decision variables) with a small probability, which adds diversity and helps avoid local optima.
    \end{itemize}

    \item \textbf{Environmental Selection}
    \begin{itemize}
        \item Combine the parent and offspring populations into a temporary pool of size \(2N\).
        \item Perform non-dominated sorting again, retaining the best \(N\) solutions (based on Pareto rank and crowding distance). These survivors form the new parent population for the next iteration.
    \end{itemize}

    \item \textbf{Iteration}
    \begin{itemize}
        \item Steps 2 through 6 repeat until reaching a \textbf{preset generation limit} or until objective improvements (e.g., the hypervolume indicator) plateau, indicating convergence.  2  through  6  repeat until reaching a \textbf{preset generation limit} or until objective improvements (e.g., the hypervolume indicator) plateau, indicating convergence.
        \item The final population typically approximates the \textbf{Pareto front}, revealing trade-off solutions among economic, environmental, and social objectives.
    \end{itemize}
\end{enumerate}

By iterating these steps, NSGA-II balances exploration across the feasible region, converging towards the Pareto-optimal set. The resulting non-dominated solutions represent diverse policy combinations that balance net revenue, environmental quality, and social satisfaction in Juneau’s tourism sector.

\section{Model Applications and Problem Solving}

\subsection{Task 1: Application in Juneau}
\label{sec:application_juneau}

In this section, we apply our integrated system dynamics and multi-objective optimization model to Juneau, demonstrating how different policy levers can influence the city's tourism sustainability. By running the NSGA-II procedure over historical and projected data (2008--2024), we obtain a set of Pareto-optimal solutions. Each solution specifies a combination of tax rate, carbon fee, capacity limits, and expenditure proportions (e.g., environment, infrastructure, community, marketing), collectively shaping the system’s long-term performance.

\begin{figure}[h]
    \centering
    \includegraphics[width=6cm]{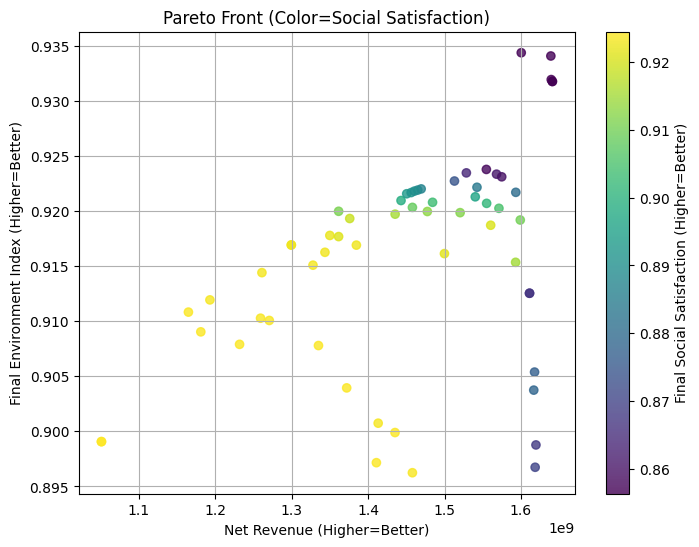}
    \caption{Pareto Front For Juneau}
    \label{fig:pareto_front}
\end{figure}

\subsubsection{Pareto Front Results.}
Figure~\ref{fig:pareto_front} illustrates the final set of non-dominated solutions in terms of \emph{(x-axis) cumulative net revenue} from 2008 to 2024 and \emph{(y-axis) final environmental index}, with bubble color indicating \emph{final social satisfaction}. The resulting trade-off curve reveals:

\begin{itemize}
    \item \textbf{High-Revenue Points.} For instance, \textit{Sol\,1/2} achieve net revenue near \(\$1.64 \times 10^9\) alongside an environmental index of 0.93. However, social satisfaction settles around 0.86. This suggests that high fiscal gain and solid environmental conditions can be realized, though local satisfaction is somewhat compromised.
    \item \textbf{Balanced Solutions.} \textit{Sol\,5} (net revenue \(\approx 1.55 \times 10^9\); environment 0.92; society 0.86) offers a more moderate balance, allowing relatively robust performance across all three objectives.
    \item \textbf{Community-Focused Options.} \textit{Sol\,0} yields a lower net revenue (\(\approx 1.46 \times 10^9\)) but pushes social satisfaction up to 0.92 and still maintains a respectable environment level of 0.90. This trade-off may be attractive when strengthening resident well-being is a top priority.
\end{itemize}

\subsubsection{Allocation of Additional Revenue and Feedback Mechanisms.}
One crucial mechanism in our model is how excess government revenue (after baseline expenses) is partitioned. These allocations directly feed back into the system by:
\begin{enumerate}
  \item \textbf{Infrastructure Investment:} Increasing carrying capacity (\(\text{capacity\_limit}\)) when additional revenue is devoted to improving ports, roads, or accommodation facilities.
  \item \textbf{Community Welfare:} Enhancing local satisfaction by funding social programs; when community satisfaction rises, residents become more welcoming to tourists, stimulating positive word-of-mouth and further revenue.
  \item \textbf{Environmental Protection:} Improving the environment index \(E_t\) through glacier-focused spending or general waste management. A healthier environment raises tourist attractiveness, promoting a virtuous cycle.
  \item \textbf{Marketing Initiatives:} Boosting visitor demand (beyond baseline projections) by advertising local attractions. This can increase tax and carbon-fee revenue, which in turn supports further sustainable investments.
\end{enumerate}
Hence, a feedback loop arises: more visitors lead to more revenue, which can be reinvested to sustain the environment or community, which then attracts still more visitors, and so on. Conversely, neglecting any aspect (e.g.\ community welfare) can degrade satisfaction and ultimately reduce visitor inflows.

\begin{figure}[h]
    \centering
    \includegraphics[width=10cm]{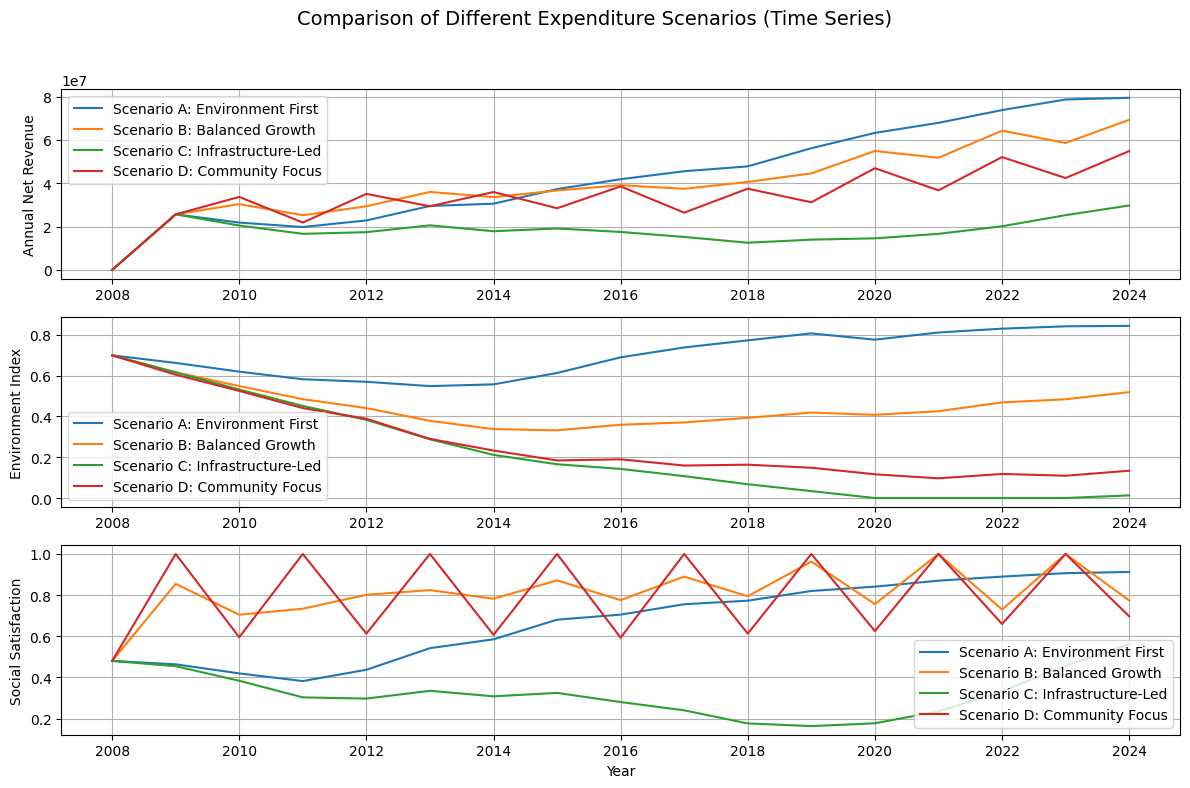}
    \caption{Scenario Comparison For Juneau}
    \label{fig:scenarioComparison}
\end{figure}

\subsubsection{Scenario Analysis for Expenditure Feedback.}
To illustrate how different expenditure priorities affect Juneau’s long-term performance, we designed four scenarios:
\begin{table}[htbp]
\centering
\begin{tabular}{l cccc}
\toprule
Scenario & $\theta_{\text{env}}$ & $\theta_{\text{infra}}$ & $\theta_{\text{community}}$ & $\theta_{\text{marketing}}$ \\
\midrule
Environment First & 0.7 & 0.1 & 0.1 & 0.2 \\
Balanced Growth & 0.3 & 0.25 & 0.25 & 0.25 \\
Infrastructure-Led & 0.3 & 0.6 & 0.1 & 0.0 \\
Community Focus & 0.2 & 0.2 & 0.6 & 0.3 \\
\bottomrule
\end{tabular}
\caption{Budget allocation scenarios and their corresponding parameter values}
\label{tab:scenarios}
\end{table}
Each scenario runs the same core system dynamics but channels excess revenue differently. Figure~\ref{fig:scenarioComparison} compares the evolution of annual net revenue, environment index, and social satisfaction from 2008 to 2024. We observe, for example, that the \emph{Environment First} scenario (blue) maintains relatively high environmental quality and eventually yields strong revenue, whereas \emph{Infrastructure-Led} (green) significantly boosts capacity yet risks greater environmental degradation. These outcomes highlight the trade-offs inherent in different allocation strategies.

\subsubsection{Sensitivity Analysis and Key Factors.}
\begin{figure}[h]
    \centering
    \includegraphics[width=14cm]{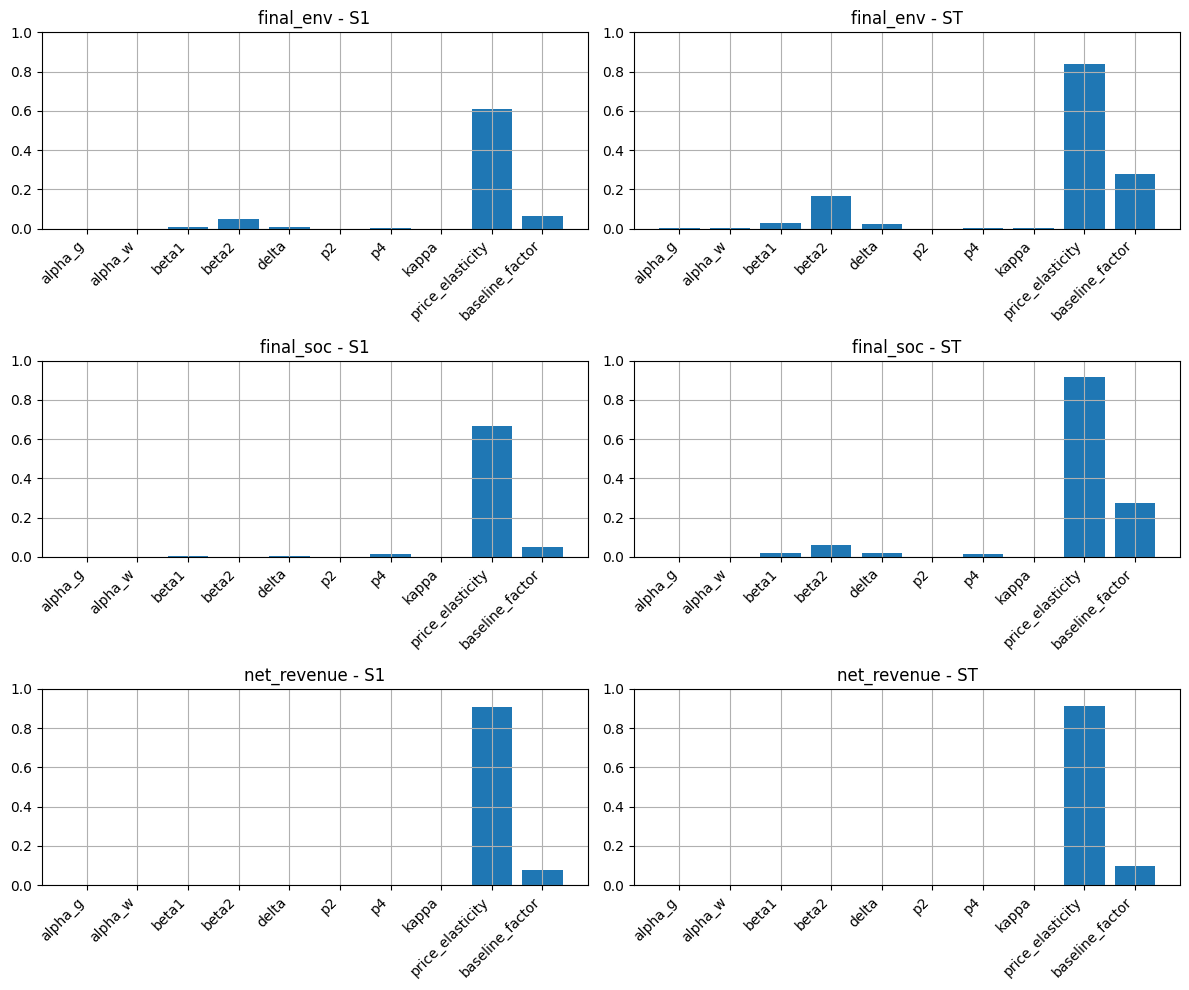}
    \caption{Sobol Final Result}
    \label{fig:Sobol_Final Result}
\end{figure}

\begin{figure}[h]
    \centering
    \includegraphics[width=15cm]{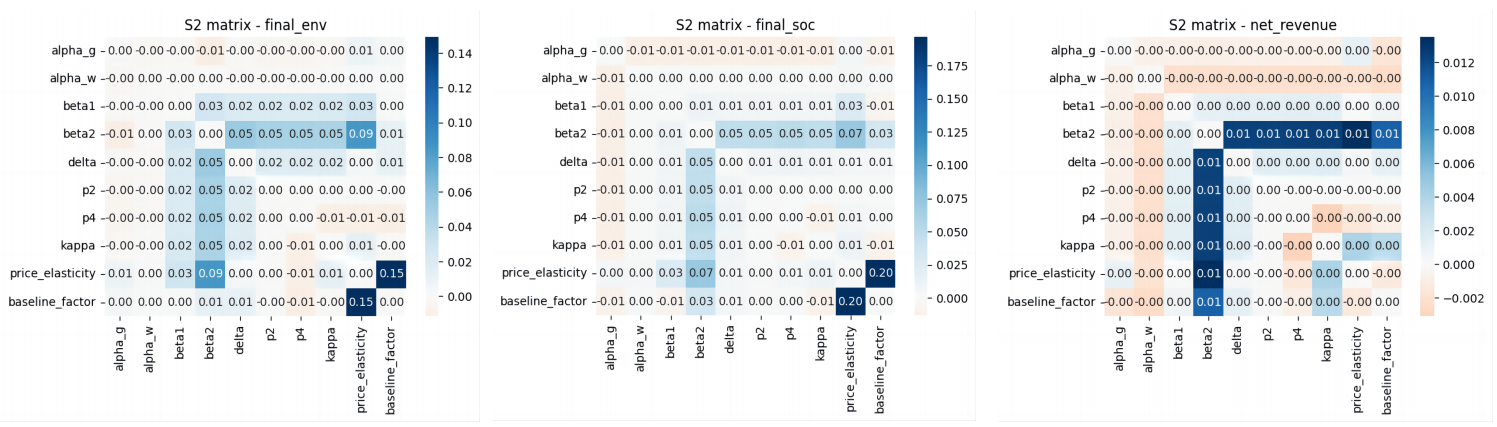}
    \caption{Sensitivity Matrices}
    \label{fig:sensitivity_matrices}
\end{figure}

We assessed global sensitivity analysis by Morris screening and Sobol analysis, a dual-method approach alternative that was complementary to the local sensitivity used on system parameters in this work evaluation. Elementary effects are used to estimate $\mu^*$ and $\sigma$ by the Morris method, where $\mu^*$ measures the influence of each parameter and $\sigma$ quantifies the parameter interactions. The first round of screening highlighted multiple parameters whose $\mu^*$ were extremely high suggesting a significant contribution towards system outputs. This in turn enabled us to extend our analysis with a Sobol sensitivity due which computed both first-order $S_i$ and total effect indices ($S_{Ti}$). First-order indices provide the amount of the direct effect of a singe parameter and total effect index include all interaction effects. Preliminary results indicate that, as shown in Figures~\ref{fig:Morris_Final Result}, \ref{fig:Sobol_Final Result} and \ref{fig:sensitivity_matrices}:
\begin{figure}[h]
    \centering
    \includegraphics[width=15cm]{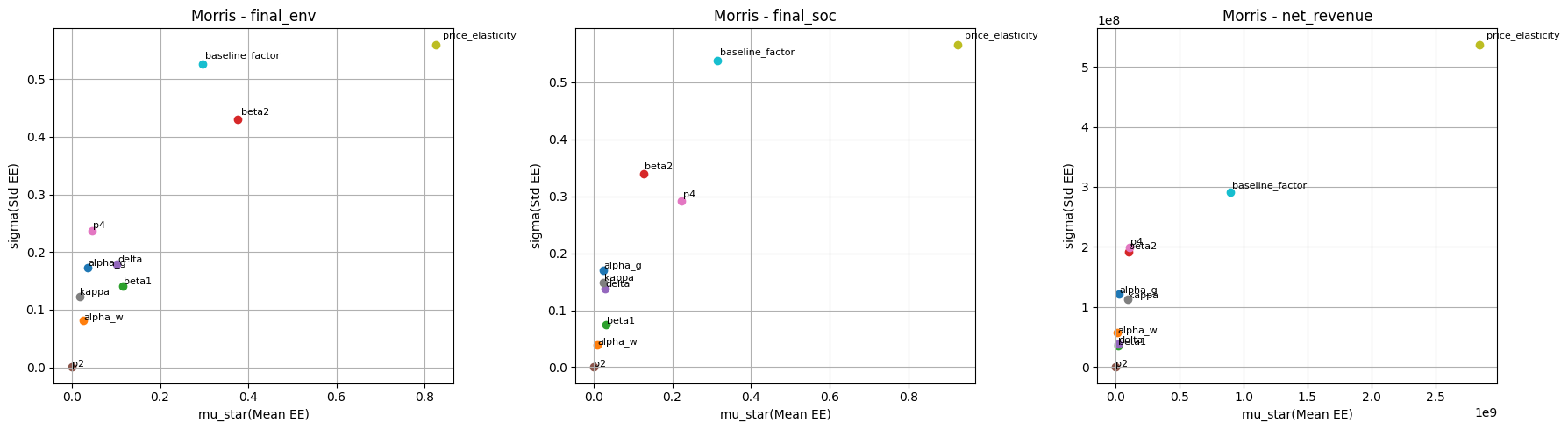}
    \caption{Morris Final Result}
    \label{fig:Morris_Final Result}
\end{figure}

\begin{itemize}
  \item \textbf{Price elasticity} of tourism demand (\(\varepsilon_{\text{price}}\)) has a pronounced impact on overall net revenue and visitor flow. Small changes in tax or carbon fees can trigger large changes in arrivals.
  \item \textbf{Environmental effectiveness coefficients} (\(\alpha_g, \alpha_w\)) and \(\beta\)-factors for glacier retreat or CO\(_2\) strongly affect \(E_T\). Overestimating or underestimating these can significantly alter the predicted environmental trajectory.
  \item \textbf{Government baseline spending fraction} also plays a major role, as it sets the pool of funds that can be reallocated to environment and community programs.
\end{itemize}

\subsection{Task 2:Application in Iceland}
For instance, to show the generality and portability of the JST model, we could also extend this to different regions, such as Iceland, where we modify the revenue and expenditure structure, environmental impact factors, and social satisfaction parameters appropriately\cite{Margo2021}. The specific model is as follows.

These analogous and contrasting characteristics not only mirror the variations in tourism development pathways between the two regions but are also essential touch points for tailoring JST (Juneau Sustainable Tourism) to its context in Iceland\cite{Petursdottir2020}. We will then adjust the IST (Iceland Sustainable Tourism) model based on our background research for Iceland.

\begin{enumerate}
    \item \textbf{Government Finance:} Adapt taxes from cruise to flight/arrival fees, include Ring Road and site fees, and calculate separate road maintenance and environmental restoration costs.

    \item \textbf{Tourist Capacity:} Consider airline capacity, Ring Road traffic, and attraction limits, with potential for dynamic pricing and a daily visitor reference capacity (\texttt{capacity\_limit}).

    \item \textbf{Environmental Evolution:} Incorporate geographical factors for glacier retreat and landscape changes. Model aviation CO2 emissions and renewable energy benefits to quantify government investments in clean energy.

    \item \textbf{Social Satisfaction:} Include a "Housing and Cost of Living Index," explore fiscal subsidies or rental controls, and consider cultural identity and local residents' tourism tolerance.
\end{enumerate}

\subsubsection{Multi-Objective Solutions for Iceland}

In order to adapt our multi-objective tourism model to Iceland’s unique circumstances, we made several targeted adjustments in both the \textbf{input data} and \textbf{model parameters}. This ensures that the simulation captures Iceland’s reliance on flight arrivals, the prominence of the Ring Road for domestic travel, and the distinctive pressures on its glaciers and local communities.

\begin{figure}[h]
    \centering
    \includegraphics[width=6cm]{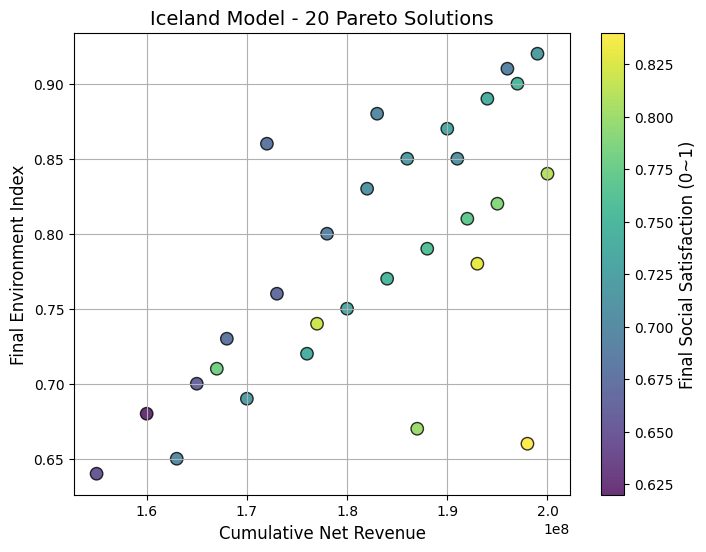}
    \caption{Pareto Front For Iceland}
    \label{fig:Pareto Front For Iceland}
\end{figure}

\begin{enumerate}
    \item \textbf{Data and Baseline Adjustments:} We replaced the Juneau dataset with Iceland-specific data for visitor arrivals, government finances, environmental indicators, and population (2008-2024). The Icelandic model considers 35\% of government expenses as tourism-related, differing from Juneau's 30\%.

    \item \textbf{Expanded Parameter Bounds:} We increased the capacity limit to \(5 \times 10^6\) and allowed a higher carbon fee (up to 120 USD per visitor), reflecting Iceland's capacity and stricter CO$_2$ policies. The ship/flight limit range was set to 500-900.

    \item \textbf{Environmental and Social Parameter Tweaks:} We adjusted glacier retreat data (200-300 feet from 2008-2024) and increased the glacier sensitivity factor (\(\kappa = 0.25\)). We also introduced a randomized initial environment index (\(0.65 \pm 0.05\)) and increased the negative crowding coefficient (\(p_2\)) to reflect housing challenges.

    \item \textbf{Infrastructure and Marketing Parameters:}  We used an  \texttt{infra\_efficiency\_factor} of 4000 and a \texttt{marketing\_efficiency\_factor} of 2500 to reflect Iceland’s geographical constraints and potential for visitor redistribution.

    \item \textbf{Evolutionary Search and Pareto Front:} We expanded the NSGA-II population to 120 for 80 generations. The resulting solutions (Figures~\ref{fig:Pareto Front For Iceland}) show trade-offs between revenue (\(1.5 \times 10^8\) to \(2.0 \times 10^8\) USD), environmental indices (up to 0.92), and social satisfaction (often reaching 0.80 or higher).
\end{enumerate}

\begin{figure}[h]
    \centering
    \includegraphics[width=10cm]{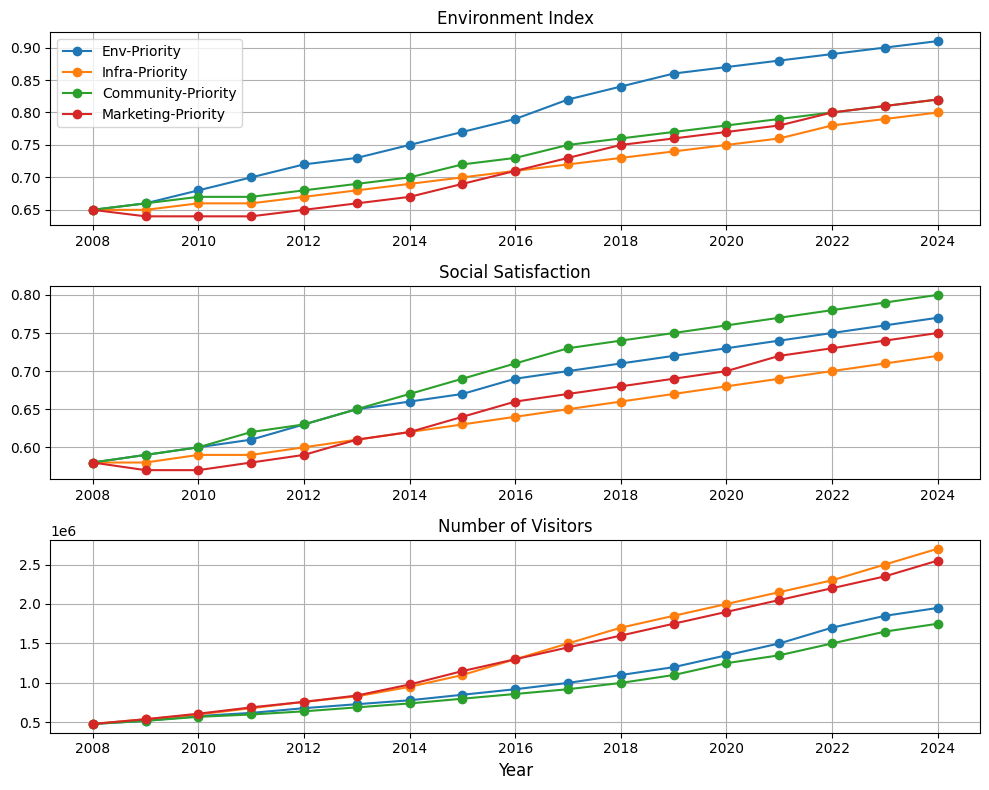}
    \caption{Scenario Comparison For Iceland}
    \label{fig:Scenario Comparison For Iceland}
\end{figure}

\subsubsection{Promoting Balanced Tourism Through Scenario Analysis}

We also implement four illustrative “expenditure scenarios”—Env-Priority, Infra-Priority, Community-Priority, and Marketing-Priority—to see how key variables (e.g., environment index, social satisfaction, visitor arrivals) evolve over time. The time-series results reveal as Figures~\ref{fig:Scenario Comparison For Iceland}:

\begin{itemize}
    \item \textbf{Env-Priority} boosts environmental resilience around glaciers and fragile landscapes, moderating visitor surge but yielding higher long-term ecological stability.
    \item \textbf{Infra-Priority} invests in roads, lodging capacity, and airport expansions, prompting faster visitor growth yet raising risks of crowding and environmental strain.
    \item \textbf{Community-Priority} channels significant funds to housing subsidies and social programs. While overall arrivals might slow, local acceptance improves noticeably, mitigating tourist backlash.
    \item \textbf{Marketing-Priority} seeks to redirect travelers from already popular sites (like the Golden Circle) to lower-traffic areas (e.g., remote waterfalls, coastal bird cliffs). This scenario often lifts satisfaction in previously overcrowded hotspots and fosters more uniform visitor distribution—albeit with potential challenges if lesser-known sites lack sufficient baseline infrastructure.
\end{itemize}

\subsubsection{Multi-Attraction Flow Distribution in Iceland}

\begin{figure}[h]
    \centering
    \includegraphics[width=12cm]{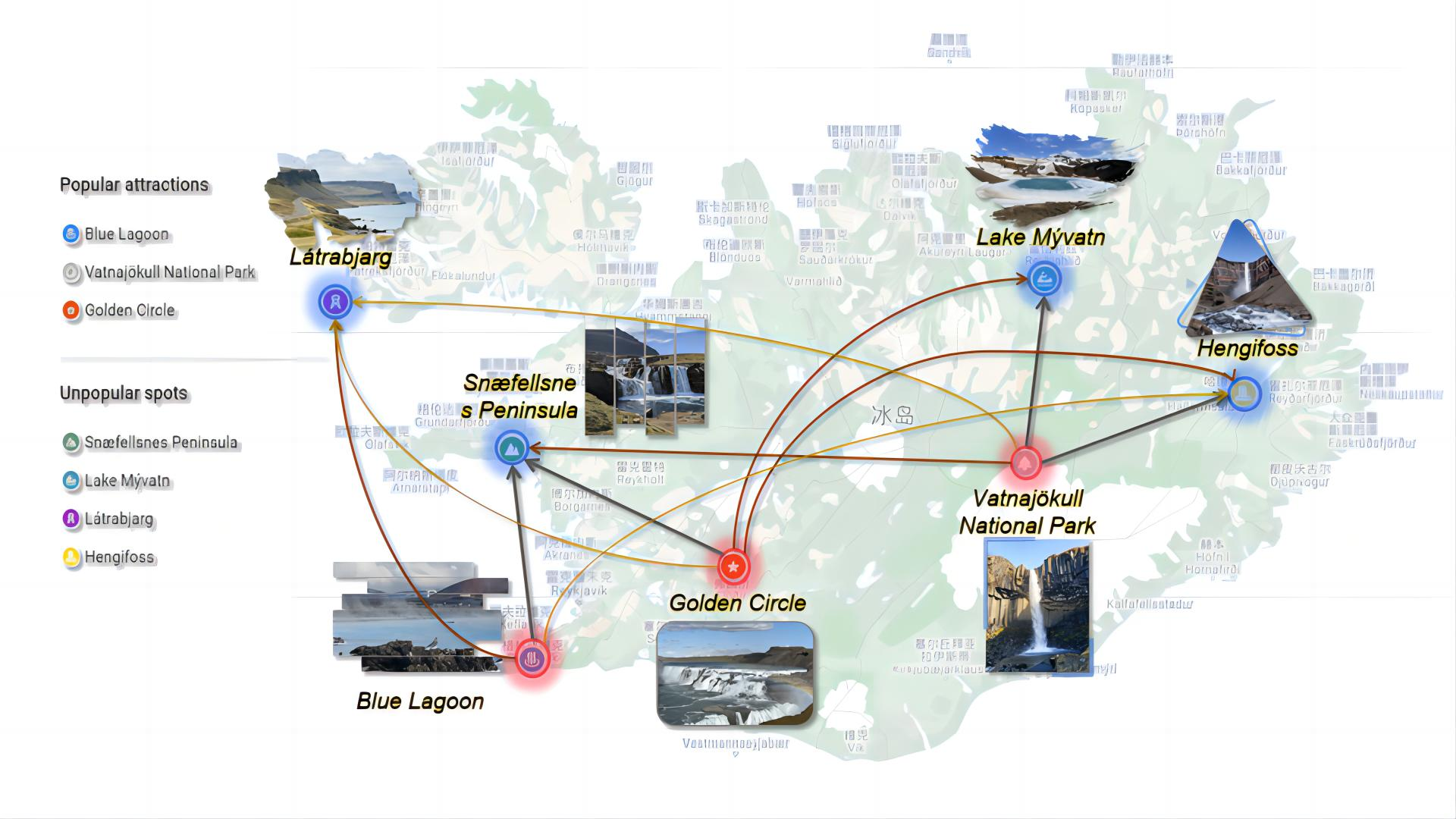}
    \caption{Tourist Flow Redistribution}
    \label{fig:Tourist Flow Redistribution}
\end{figure}

As illustrated in Figures~\ref{fig:Tourist Flow Redistribution}, to address the uneven distribution of tourists across Iceland’s attractions—where destinations such as the Blue Lagoon, Vatnajökull, and the Golden Circle have become overcrowded, while equally remarkable sites like Snæfellsnes, Mývatn, Látrabjarg, and Hengifoss remain underutilized—we have developed a multi-attraction diversion model. By strategically adjusting key decision variables such as marketing intensity, pricing strategies, and infrastructure development, this model seeks to promote a more balanced visitor flow and thereby foster sustainable development in Iceland.

\textbf{Core Equations for Annual Redistribution}

Our Multi-Attraction Flow Distribution iterates yearly from \(t\) to \(t+1\). The following steps summarize the logic:

\begin{enumerate}
    \item \textbf{Island-wide Potential Visitors} \(\displaystyle T(t)\):
        \begin{equation}
        T(t+1) \;=\; T(t)\,\times \Bigl[1 \;+\; \phi \,\bigl(\text{Price}_{\text{avg}}(t)+\text{Env}_{\text{avg}}(t)-1.5\bigr)\Bigr]
        \;+\;\mathrm{DevBoost}.
        \end{equation}
         The total island-wide potential visitors \(T(t+1)\) is calculated based on the previous year's visitors \(T(t)\), adjusted by a factor that incorporates the average price and environmental conditions (\(\text{Price}_{\text{avg}}(t)\) and \(\text{Env}_{\text{avg}}(t)\)) across all sites, and a development boost (\(\mathrm{DevBoost}\)) to account for government-driven campaigns.

    \item \textbf{Site Attractiveness} \(A_i(t)\):
         \begin{equation}
        A_i(t) \;=\;
        \exp\Bigl[\alpha_{0} \;+\; \alpha_{1}\,E_i(t) \;+\; \alpha_{2}\,S_i(t)
        \;+\; \alpha_{3}\,\ln\bigl(1 + \mathrm{Mkt}_i(t)\bigr)
        \;-\; \alpha_{4}\,\mathrm{Price}_i(t)\Bigr].
        \end{equation}
        Each site's attractiveness \(A_i(t)\) is determined by an exponential function that considers the site's environmental index \(E_i(t)\), social satisfaction \(S_i(t)\), marketing effort \(\mathrm{Mkt}_i(t)\), and price level \(\mathrm{Price}_i(t)\). These factors are weighted by coefficients \( \alpha_{0},\alpha_{1},\dots,\alpha_{4}\) that capture their relative importance.

    \item \textbf{Preliminary Visitor Count} \(V_i^{\mathrm{raw}}(t+1)\):
         \begin{equation}
        V_i^{\mathrm{raw}}(t+1) \;=\; p_i(t)\;\times\;T(t+1).
        \end{equation}
        The preliminary visitor count \(V_i^{\mathrm{raw}}(t+1)\) for each site is calculated as the product of the allocation weight \(p_i(t)\) and the total island-wide potential visitors \(T(t+1)\).

    \item \textbf{Environment Update}:
         \begin{equation}
        E_i(t+1)
        \;=\;
        E_i(t)
        \;+\;
        \alpha_g\,\mathrm{EnvFund}_i(t)
        \;-\;
        \beta_{\mathrm{crowd}}
        \,\frac{V_i(t)}{C_i(t)}
        \;-\;
        \beta_{\mathrm{CO2}}\;\mathrm{CO2}_i(t)
        \;+\;
        \delta\,[1 - E_i(t)].
        \end{equation}
       The environment index \(E_i(t+1)\) for each site is updated based on the previous year's index \(E_i(t)\), enhanced by site-specific environmental funding (\(\mathrm{EnvFund}_i(t)\)), reduced by crowding (\(\beta_{\mathrm{crowd}} \,\frac{V_i(t)}{C_i(t)}\)) and CO$_2$ emissions (\(\beta_{\mathrm{CO2}}\;\mathrm{CO2}_i(t)\)), and increased by natural restoration (\(\delta\,[1 - E_i(t)]\)).

    \item \textbf{Social Satisfaction Update}:
         \begin{equation}
        S_i(t+1) \;=\;
        S_i(t)
        \;+\;
        \rho_{\mathrm{comm}}\;\mathrm{CommunityFund}_i(t)
        \;-\;\rho_{\mathrm{over}}\,
        \dfrac{V_i(t)}{\mathrm{Pop}_i}
        \;+\;\rho_e\,[E_i(t+1) - S_i(t)],
        \end{equation}
        The social satisfaction \(S_i(t+1)\) for each site is updated based on the previous year's satisfaction \(S_i(t)\), boosted by community-specific funding (\(\mathrm{CommunityFund}_i(t)\)), reduced by the visitor-to-population ratio (\(\dfrac{V_i(t)}{\mathrm{Pop}_i}\)), and adjusted by the new environment index (\(E_i(t+1)\)).
\end{enumerate}

\begin{figure}[h]
    \centering
    \includegraphics[width=10cm]{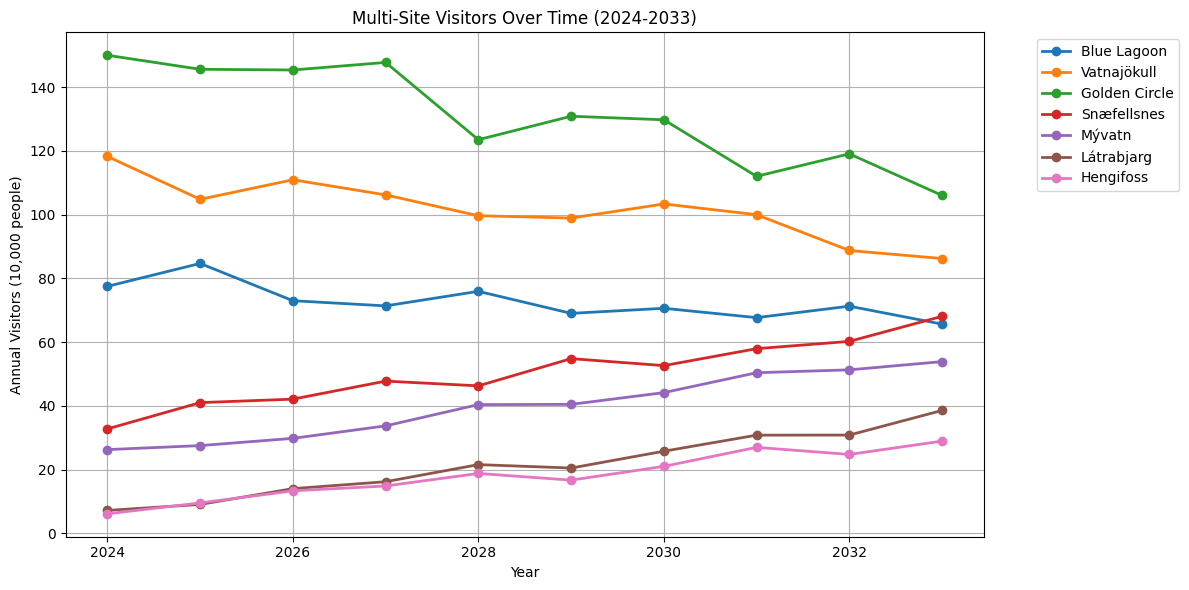}
    \caption{Time Series of Tourist Arrivals}
    \label{fig:Time Series of Tourist Arrivals}
\end{figure}

\begin{figure}[h]
    \centering
    \includegraphics[width=8cm]{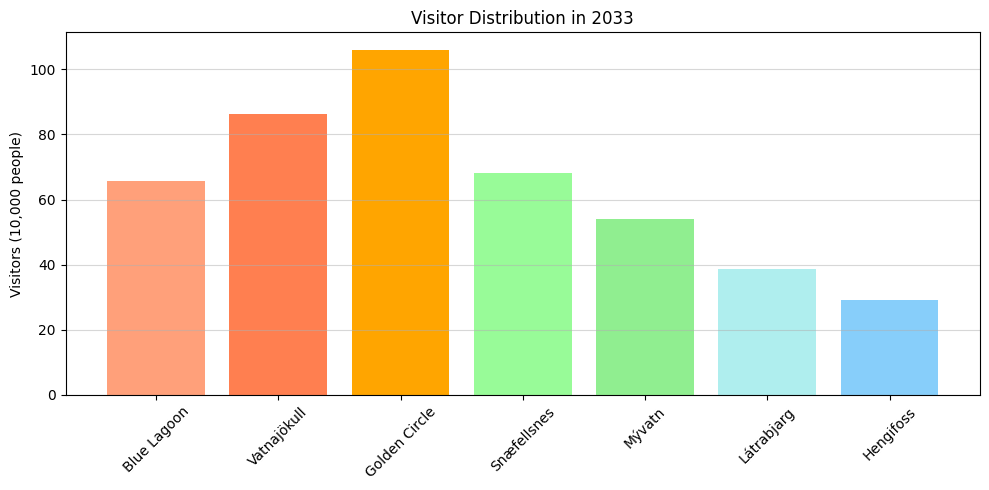}
    \caption{Visitor Distribution in Attractions}
    \label{fig:Visitor Distribution in Attractions}
\end{figure}

\subsubsection{Visualizing the Redistribution and Final Distribution}

Figures~\ref{fig:Time Series of Tourist Arrivals} illustrates a policy scenario (2024-2033) that shifts marketing efforts to lesser-known sites (e.g., Látrabjarg, Hengifoss). This results in a downward trend in visitor counts at hotspots (Blue Lagoon, Vatnajökull, Golden Circle), while lesser-visited sites experience a steady increase, though remaining below the original hotspots' levels.

Figures~\ref{fig:Visitor Distribution in Attractions}  shows the final visitor distribution in 2033, with previously overlooked areas absorbing a greater share of total visits. This more evenly spread tourism landscape, supported by infrastructure investments and local engagement, balances economic benefits across multiple regions and reduces overexploitation of key attractions.

\section{Discussion and Limitations}
\subsection{Discussion}
\begin{itemize}
    \item \textbf{Holistic Perspective}: Integrates economic, environmental, and social outcomes to capture the true breadth of sustainable tourism challenges.
    \item \textbf{Adaptable Framework}: Can be recalibrated to different destinations by tuning parameters for capacity, carbon fees, and community responses.
    \item \textbf{Decision-Support Depth}: Offers both Pareto-optimization and scenario-based analysis, clarifying policy trade-offs and enabling robust planning.
\end{itemize}

\subsection{Limitations and Future Work}
\begin{itemize}
    \item \textbf{Data Sensitivity}: Accuracy depends on the reliability of visitor demand, environmental impact, and social tolerance parameters, which may be difficult to estimate.
    \item \textbf{Simplifying Assumptions}: Cannot fully replicate complex or rapidly shifting real-world conditions (e.g., extreme events, evolving traveler behaviors).
    \item \textbf{Future Work}: Future extensions could incorporate machine-learning-based demand forecasting modules, richer behavioral heterogeneity for tourists and residents, and more detailed policy instruments (e.g., dynamic pricing or adaptive tax schemes) to further bridge system-dynamics modeling with modern AI-enabled decision support.
\end{itemize}

\end{document}